\documentclass[preprint,review,12pt,compress]{elsarticle}

\usepackage{graphicx}
\usepackage{epsfig}
\usepackage{bm}
\usepackage [T1]{fontenc}
\usepackage{color}
\usepackage{wrapfig}
\usepackage{makeidx}
\usepackage[latin1]{inputenc}
\usepackage{graphics}
\usepackage{psfrag}
\usepackage{float}

\usepackage{amssymb}

\journal{ArXiv}

\begin{document}

\begin{frontmatter}



\title{Perturbation theory for very long-range potentials}


\author{L. Escamilla\corref{cor1}}
\ead{lescamilla@fisica.ugto.mx}
\author{J. Torres-Arenas}
\ead{jtorres@fisica.ugto.mx}
\author{A. L. Benavides}
\ead{alb@fisica.ugto.mx}

\cortext[cor1]{Corresponding author. Tel: +52 477 7885100 ext.8476 Fax: +52 477 7885100 ext.8410}

\address{Divisi\'on de Ciencias e Ingenier\'ias Campus Le\'on, Universidad de Guanajuato, Loma del Bosque 103, 37150  Le\'on Guanajuato, M\'exico.}

\begin{abstract}
Systems with very long-range interactions (that decay at large distances like $U(r)\sim r^{-l}$ with $l\le d$ where $d$ is the space
dimensionality) are difficult to study by conventional statistical mechanics. Examples of these systems are gravitational and charged (non-electroneutral). In this work we propose two alternative methodologies to avoid these difficulties and capture some of the properties of the original potential. The first one consists in expressing the original potential in terms of a finite sum of hard-core Yukawa potentials. In the second one, the potential is rewritten as a damped potential, using a damping function with a parameter that controls the range of the interaction. These new
potentials with finite ranges, which mimic the original one, can now be treated by conventional statistical mechanics methods.  
\end{abstract}


\end{frontmatter}


\section{Introduction}
\label{Intro}

Description of systems interacting via the so-called long-range interactions (LRI) is an important statistical mechanics problem. These systems are found  from very small to very large scales, for instance, in astrophysics~\cite{Padmanabhan,Vega,Chavanis}, plasma physics~\cite{Kiessling,Lebowitz},
hidrodynamics~\cite{Miller}, atomic physics~\cite{Schmidt} and, nuclear physics~\cite{D'Agostino}.

In order to use a precise definition of LRI between a pair of particles, that are a distance $r$ apart, we consider the following: when the interaction potential between particles decays at long distances like $1/r^{l}$ in a space of $d$ dimensions, the interaction can be considered to be long-range if $l \le d$.\\
This definition is a consequence of considering  the energy $e$ of a given particle located at the center of a sphere of radius $R$ with a homogeneous particle distribution in $d$-dimensions. In order to exclude the divergence that appears at very short distances,  the  energy of the neighboring particles located inside a sphere of radius $\delta$ is neglected, $e$ is given by,

\begin{displaymath}
e =\int_\delta^R\frac{\rho B}{r^l}d^dr=\rho B\Omega_d\int_\delta^R r^{(d-1)-l}dr
\end{displaymath}
\begin{equation}\label{eq1}
\quad =\frac{\rho B\Omega_d}{d-a}[R^{d-l}-\delta^{d-l}]; \qquad \textrm{if} \ l\neq d;
\end{equation}
where $\rho$ is the generic particle density, $B$ is a coupling
constant which guarantees the correct energy dimensions, and
$\Omega_d$ is the angular volume in the $d$-dimensional space. When
$R$ is increased, $e$ remains finite only when $l>d$; such cases are
the usual short-range interactions. The opposite corresponds to $l\le
d$, where energy diverges for an increasing volume; these are
long-range interactions. Examples of different long-range potentials
are shown in Figure 1. Notice that this particular definition could be different in the context of fluids theory.

In statistical mechanics, most of the effort to obtain the equilibrium and non-equilibrium thermodynamic properties, has been concentrated on
systems with short-range interactions. One of the main features of LRI systems is that their total energy, under the pairwise additive approximation, is non-extensive, and as a consequence, is also non-additive~\cite{Tsallis,Campa,Dauxois,Kac,Oppenheim,Gross}. Therefore, the connection between Boltzmann-Gibbs (BG) statistical mechanics and classical thermodynamics is not straightforward, since the latter assumes that energy is an additive quantity~\cite{Callen}. To our knowledge, there is not a thermodynamic formalism (independent of a statistical mechanics approach) that allows this connection; however, it is possible to start from a non-extensive statistical mechanics and to obtain a non-extensive thermodynamic formalism. \\ 
Perhaps a non-extensive version of statistical mechanics could be a more natural theoretical frame to study LRI.  A few proposals for BG statistical mechanics generalizations have been given~\cite{Tsallis,Renyi,Rajagopal}, however none of them are unanimously accepted. Besides, the application of these
generalizations to long-range potentials has been scarce. Another approach is to make adjustments to the BG formalism to study these systems~\cite{Vega,Campa,Saslaw}.

In order to avoid the difficulties to treat LRI mentioned above, in this work we present a first naive approach, but general, in the sense that it can be applied to a great variety of long-range potentials in the frame of BG statistical mechanics. This methodology consists  in rewritting a long-range potential as a short- range one, being the latter  similar to the long-range potential in its graphical representation. We expect that this short-range potential recovers some features of the original one, however we know that this path leads to a classical thermodynamics frame and we do not know if real systems with LRI are well represented by this thermodynamics.\\
More specifically our approach consists in expressing a given long-range potential as a) a finite sum of Yukawa potentials; b) a product of the potential with a damping function which depends on a parameter, that under a certain  limit, the original potential is recovered. \\
We have selected the  Discrete Perturbation Theory~\cite{Benavides} (DPT) and the First-order Mean Spherical Approximation~\cite{Tang} (FMSA) to study these potentials. These theories have been successfully applied in the context of fluids and more recently in the soft matter field, and can be applied to a great variety of potentials. 
 
As an illustrative example, we have choosed the gravitational potential due to the interaction between two identical spherical rotating bodies (ETS potential), obtained by Escamilla et al.~\cite{Escamilla}. More interesting models in the context of molecular liquids,  could be, for instance,  the Coulomb interactions.

This work is organized as follows. In section \ref{theo}, we give a brief description of the ETS potential  and of the hard-core multi-Yukawa (HCMY) and damped potential approaches. In section \ref{res} we present internal energies, pressures, and vapor-liquid phase diagrams for the approximated potential. Finally, in section \ref{concl} we give the main conclusions of this work.

\section{Theory}
\label{theo}

\subsection{The ETS potential}

In the context of general relativity, within the weak-field limit methodology,  an angular averaged  potential due to the interaction between two identical spherical rotating bodies was proposed~\cite{Escamilla}. This interaction potential for hard-core spheres is given by $U^*(x)\equiv U(x)/\mid\epsilon_{min}\mid$, where:

\begin{equation}\label{eq2}
U^*(x) =\left\{ \begin{array}{ll}
\infty & \textrm{if} \ x < 1 \\
\\ -\frac{1}{\arctan(\alpha^*)}\arctan\left(\alpha^*/x\right) & \textrm{if} \ x \ge 1;
\end{array} \right.
\end{equation}
with  $\alpha^*\equiv J/Mc\sigma$, $M$ is the mass, $J$ is the angular momentum, $c$ is the speed of light in vacuum, $\sigma$ is the diameter of particles, $x=r/\sigma$ and $\epsilon_{min}$ is the potential evaluated at $x=1$. 
This potential is purely attractive, nondivergent at short distances (for $x \to 0$ , $1/\alpha^*\arctan(\alpha^*/x)\to\pi/2$), and keeps its long-range nature satisfying the condition given by (\ref{eq1}). Specific angular momentum $\alpha^*$ is a parameter which modulates the intensity of the interaction. The long-range behavior is the same for any finite  value of $\alpha^*$. For instance, in the limit of $\alpha^* \to0$ the ETS potential goes to conventional  $-1/x$ Newtonian gravitational interaction. In Figure~2, the ETS and $-1/x$ potentials are shown; it can be noticed that the long-range behavior is the same for both of them. 

To avoid the difficulties to evaluate thermodynamic properties for this long-range potential, we propose to rewrite the non hard-core potential part as:\\ 
1) a finite sum of $m$ Yukawa potentials,

\begin{equation}\label{eq3}
\Phi_{MY}(x)=\sum_{i=1}^m=\epsilon_i^*\frac{\exp[{-\kappa_i^*(x-1)}]}{x}; \quad \textrm{\textit{multi-Yukawa approach},}
\end{equation}
with the energy and range parameters $\epsilon_i^*=\epsilon_i/\mid \epsilon_{min}\mid$ and $\kappa_i^*$, respectively, and\\
2) a damped potential, which consists in the product of the original potential and a damping function $f(\gamma,x)$,

\begin{equation}\label{eq4}
\Phi_D(\gamma,x)=f(\gamma,x)\frac{U(x)}{\mid\epsilon_{min}\mid}; \qquad \quad \textrm{\textit{damped approach},}
\end{equation}
where $\gamma$ is the damping parameter that can be selected in order to guarantee that the approximated potential mimics the original one.

\subsection{First-order mean spherical aproximation}

The first-order mean spherical approximation was developed by Tang et al.,~\cite{Tang2} as an improvement of the mean spherical aproximation (MSA)~\cite{Waisman}. 

The solution of the Ornstein--Zernike integral equation under MSA makes it possible to  find analytical thermodynamic and structure expressions, which otherwise would require time-consuming numerical work. Despite these advantages, MSA may lose its solution in unstable regions~\cite{Tang3}. An improvement of this theory is the first-order mean spherical approximation.\\
FMSA solves analytically the radial distribution function (RDF)  to first order in terms of inverse temperature.  Solutions obtained are explicit, simpler and always exist in unstable regions~\cite{Tang3}. 

A successful  application of FMSA theory was done by Tang et al.,~\cite{Tang2} to the Yukawa and HCMY potentials. A finite sum of Yukawa potentials can mimic other well-known potentials, like Lennard-Jones potential~\cite{Tang} or sticky hard spheres~\cite{Baxter}; however, its efficacy for LRI like the ETS potential has never been tested to our knowledge. In this work we will approximate the ETS potential using HCMY potential to express it.
 
The hard-core Yukawa potential with multiple tails $m$, for hard-core particles is given by,

\begin{equation}\label{eq5}
\Phi_{MY}(x)=\left\{ \begin{array}{ll}
\infty & \textrm{if} \ x  <  \sigma \\
\\ \sum_{i=1}^{m}\epsilon_i^*\frac{\exp{[-\kappa^*_i(x-1)]}}{x} & \textrm{if} \ x \ge \sigma,
\end{array} \right.
\end{equation}

For a system of $N$ particles confined in a volume $V$ at temperature $T$ interacting through a HCMY potential (\ref{eq5}), the reduced  Helmholtz free energy ($a = A/Nk_BT$, $k_B$ is the Boltzmann's constant), within the FMSA can be expressed as~\cite{Tang},

\begin{equation}\label{eq6}
a=a_{ideal}(\eta,T)+a_{hs}(\eta) + a_1 (\eta, T, \kappa^*_i, \epsilon^*_i) + a_2 ((\eta, T, \kappa^*_i, \epsilon^*_i);
\end{equation}
where $a_{ideal}(\eta,T)$ and $a_{hs}(\eta)$ are the ideal gas and hard-sphere contributions to free energy, $\eta=\pi\rho^*/6$ is the packing fraction and $\rho^*=\rho\sigma^3$ is the reduced density. The Carnahan and Starling EOS~\cite{Carnahan} for $a_{hs}(\eta)$ is used,

\begin{equation}\label{eq7}
a_{hs} (\eta) =\frac{4\eta-3\eta^2}{(1-\eta)^2}.
\end{equation}
The first and second order perturbation terms,  $a_1(\eta, T, \kappa^*_n, \epsilon^*_n)$ and $a_2(\eta, T, \kappa^*_n, \epsilon^*_n)$ are defined as

\begin{equation}\label{eq8}
a_1(\eta, T, \kappa^*_n, \epsilon^*_n)=12\eta\beta\sum_{i=1}^{m}\frac{\epsilon^*_i L(\kappa^*_i)}{(1-\eta)^2 Q_0(\kappa^*_i)(\kappa^*_i)^2},
\end{equation}
\begin{equation}\label{eq9}
a_2(\eta, T, \kappa^*_n, \epsilon^*_n)=6\eta\beta^2\sum_{i=1}^m\sum_{j=1}^m\frac{\epsilon^*_i\epsilon^*_j}{(\kappa^*_i+\kappa^*_j)Q^2_0(\kappa^*_i)Q^2_0(\kappa^*_j)};
\end{equation}
with $\beta=1/k_BT$, $n = 1, \dots , m$. The other functions that appear in (\ref{eq8}) and (\ref{eq9}) are defined by

\begin{displaymath}
Q_0(t)=\frac{S(t)+12\eta L(t)e^{-t}}{(1-\eta)^2t^3},
\end{displaymath}
\begin{displaymath}
S(t)=(1-\eta)^2t^3+6\eta(1-\eta)t^2+18\eta^2t-12\eta(1+2\eta),
\end{displaymath}
\begin{equation}\label{eq10}
L(t)=\Bigg(1+\frac{\eta}{2}\Bigg)t+1+2\eta.
\end{equation}
Other thermodynamic properties can be derived from (\ref{eq6}).

\subsection{Discrete perturbation theory}

The discrete perturbation theory was developed by Benavides and Gil-Villegas~\cite{Benavides} and has been successfully applied to different discrete and continuous potentials~\cite{Vidales,Cervantes,Benavides2}. In particular Torres-Arenas et al.~\cite{Torres}, have found that this approach accurately describes the thermodynamics of a hard-core attractive Yukawa potential with $\kappa^* = 0.1$, which represents a long-range potential in the context of molecular liquids; however, it is a short-range interaction in the sense of (\ref{eq1}). In this work, DPT will be used for even smaller values of $\kappa^*$. 

Analogous to FMSA, DPT provides a recipe for the Helmholtz free-energy of an arbitrary radial potential, re-expressing it as a sum of square-well and/or square-shoulder potentials.
For a system of $N$ particles, contained in a volume $V$ at temperature $T$, the reduced free Helmholtz energy is given by,

\begin{displaymath}
a=a_{hs}(\eta)+\beta\sum_{i=1}^n\Bigg[a_1^S(\eta,\lambda_i,\epsilon_i)-a_1^S(\eta,\lambda_{i-1},\epsilon_i) \Bigg]
\end{displaymath}
\begin{equation}\label{eq11}
\qquad +\beta^2\sum_{i=1}^n\Bigg[a_2^S(\eta,\lambda_i,\epsilon_i)-a_2^S(\eta,\lambda_{i-1},\epsilon_i)\Bigg];
\end{equation}
where $n$ is the total number of steps, $\epsilon_i$ is an energy parameter that can be positive or negative and defines the height/depth of the step, and $\lambda_i-\lambda_{i-1}$ is the width  of $i$-th step. $a^S_{1}(\eta,\lambda_n,\epsilon_i)$ and $a^S_{2}(\eta,\lambda_n,\epsilon_i)$ are the first-order and second-order perturbation contributions for a square-well ($\epsilon_i>0$) or square-shoulder ($\epsilon_i<0$). The evaluation of (\ref{eq11}) can be simplified since square-well (SW) and square-shoulder (SS) potentials differ only in the sign of $\epsilon_i$ and the following relations are satisfied:

\begin{displaymath}
a_1^{SS}(\eta,\lambda_i,\epsilon_i)=-a_1^{SW}(\eta,\lambda_i,\epsilon_i); 
\end{displaymath}
\begin{equation}\label{eq12}
a_2^{SS}(\eta,\lambda_i,\epsilon_i)=a_2^{SW}(\eta,\lambda_i,\epsilon_i).
\end{equation}
The problem is now reduced to calculate the free energy perturbation terms of the square-well potential.

In order to apply the DPT to a continous potential, a discrete version of the potential is required~\cite{Torres}. We used an approximation of the desired continous potential (either the HCMY or the damped potential approach) in terms of step functions, choosing the middle point on each step to evaluate the potential.\\
For monotonic potentials without a defined range, DPT requires a cutoff; this cutoff can be calculated by the condition $|U(\lambda_c)|=10^{-6}$, where $\lambda_c$ is the cutoff distance. This condition was successfully applied to Yukawa potential~\cite{Torres}.

In the DPT approach, the perturbation terms $a_1^{SW}(\eta,\lambda_i,\epsilon_i)$ and $a_2^{SW}(\eta,\lambda_i,\epsilon_i)$, have been calculated from two theoretical equations: for $1.1<\lambda\le3.0$, we used Esp\'indola et al.,~\cite{Espindola} and, for longer SW ranges, Benavides and del R\'io expressions~\cite{Benavides3}. 
DPT requires the SW equation of state for all values of $\lambda$; however, the EOS proposed by Esp\'indola et al., have problems for $\lambda<1.1$. Therefore, in this work the number of discretizations, $n$, is constrained to prevent an evaluation of the perturbation contributions for free energy $a_1^{SW}$ and $a_2^{SW}$ in $1.0<\lambda<1.1$~\cite{Torres}, $n$ is determined by $n=(\lambda_c-\lambda_0)/0.1=10(\lambda_c-\lambda_0)$.

\section{Results}
\label{res}
\subsection{Multi-Yukawa approach}

The first proposal to study the thermodynamic properties of a long-range potential is to rewrite it as a finite sum of Yukawa potentials. This goal can be achieved using the Levenberg-Marquardt method to perform a non-linear least squares fit~\cite{Gershenfeld}.\\
For the ETS potential, we have found  that three Yukawa potentials are enough to describe it (see Figure~3). Notice that for both potentials, the long-range behavior is essentially the same and differences can be seen only at short distances. We expect that these short-range dicrepancies will be irrelevant for this type of potentials. The fitting was performed for the particular case of $\alpha^*=10$.

The ETS potential in the HCMY approach is given by (\ref{eq5}) for $m=3$. Corresponding parameters $\epsilon_i^*$ and $\kappa^*_i$ are presented in Table \ref{table1}.

\begin{table}[tbph]
\caption{Strength and range parameters for HCMY approach to the ETS potential.}
\label{table1}
\centering
\par 
\begin{tabular}{ccc}$i$ & $\epsilon_i^*$ & $\kappa^*_i$ \\ \hline
1 ~&~ 5.6590 ~&~ 0.1500 \\
2 ~&~ -2341.2060  ~&~ 3.6744$\times 10^{-4}$\\
3 ~&~ 2334.5470 ~&~ 3.6880 $\times 10^{-4}$\\
\hline \hline
\end{tabular}
\end{table}

As can be seen, $\kappa_i^*$ values are small, corresponding to
long-range molecular potentials; nevertheless, they are short-ranged
if we consider definition given in (\ref{eq1}). Therefore, HCMY
potential can be treated by the conventional methods of BG Statistical Mechanics. \\
In the frame of a linear theory, as is the case of FMSA, the problem is reduced to the knowledge of an EOS for the hard-core attractive/repulsive Yukawa potential of variable $\kappa^*$.  Yukawa EOS have not been tested  for such small  $\kappa^*$ values. Results obtained in this work can also be used as a test of the reliability of FMSA and DPT in a domain of $\kappa_i\approx10^{-4}$, far from the $\kappa^*_i$ values commonly studied by liquid state community~\cite{Tang2,Torres,Mi}.\\
The reduced units that will be used are given by,

\begin{equation}\label{eq13}
T^*=\frac{k_BT}{\mid\epsilon_{min} \mid}; \ P^*=\frac{P\sigma^3}{\mid\epsilon_{min} \mid}; \ \mu^*=\frac{\mu}{k_BT}; \ \rho^*=\rho\sigma^3,
\end{equation}
where $T^*$, $P^*$, $\mu^*$, and $\rho^*$ are the reduced temperature, pressure, chemical potential, and density, respectively.

To estimate the accuracy of FMSA, compared with DPT for a hard-core attractive Yukawa potential (HCAY) with small values of the range parameter, we considered the case $\kappa^*=0.1$. For this value, DPT gives very good predictions\cite{Torres}. To our knowledge, it is the smallest value considered in simulation studies~\cite{Caillol}.\\ 
The results for a vapor-liquid phase diagram with FMSA, DPT and simulation are presented in Figure~4. Both theories are in good agreement with simulation data.

With the performance test of FMSA and DPT done, we constructed vapor-liquid phase diagrams of the HCMY approach to the ETS potential for $\alpha^*=10$. 
In Figure~5, vapor-liquid phase diagrams  for both theories are presented.\\
In the absence of simulation data for the thermodynamic properties of the ETS potential, we present only a theoretical prediction of pressure and excess internal energy as a function of density, for several isotherms.

The cutoff ($\lambda_c$) for the potentials used within DPT has been selected as $|U(\lambda_c)|=10^{-8}$ that guarantees that the thermodynamic properties do not change significatively if the cutoff is increased.

In Figure~6 and in Figure~7 we present reduced pressure and excess internal energy as a functions of density, for the approximated predictions of FMSA and DPT theories of HCMY. 
In Figure~6 the selected isotherms were chosen around the critical temperature.  For the case $\alpha^*=10$, coexistence isotherms (showing a van der Waals loop) appear at very high temperatures; which is a typical behavior for long-range potentials~\cite{Torres,Caillol}.\\
The corresponding excess internal energies shown in Figure~7,  present the same ordinary functionality as the one exhibited by a short-range potential.

\subsection{Damped potential approach}

In order to illustrate the methodology given in (\ref{eq4}) for the ETS potential, we have selected the damping function, $f(\gamma,x)=\exp[-\gamma(x-1)]$. Therefore the damped potential for $x\ge1$, can be written as:

\begin{equation}\label{eq14}
\Phi_D(\gamma,x)=-\frac{\exp[-\gamma(x-1)]}{\arctan(\alpha^*)}\arctan\Bigg(\frac{\alpha^*}{x}\Bigg);
\end{equation}
where we considered the case $\alpha^*=10$. The parameter $\gamma=1.9603\times10^{-4}$ was estimated by solving,

\begin{equation}\label{eq15}
\int_1^\infty x^2\Phi_D(\gamma,x)dx=\int_1^\infty x^2\Phi_{MY}(x)dx.
\end{equation}
This equation guarantees that, in this approach, long-range first-order perturbation term ($g(x)\approx1$, $g(x)$ being the radial distribution function) is equivalent to the one in the HCMY approach. In this way we avoided the divergence of this term, if the ETS potential is used. It is important to remark that the previous condition is not the only possible one; for instance, another potential that mimics the ETS potential expressed in a different basis, could have been chosen instead of $\Phi_{MY}(x)$.

With this $\gamma$ selection, in Figure~8 we show that damped and the ETS potentials are very similar.
This new potential will be studied using DPT. Since there are no available simulation data for the ETS potential, in the following figures we make a comparison between the two methodologies presented in this work.

In Figure~9 the  vapor-liquid phase diagrams  of  damped potential and HCMY approximations to the ETS potential are shown.  As can be seen they give very similar predictions.\\
In Figure~10 and Figure~11, reduced pressures and excess internal energies as a function of density for  damped potential and HCMY approaches are presented. It can be noticed that both methodologies are equivalent. 

\section{Conclusions}
\label{concl}

In this work two different methodologies to study long-range
potentials fluids from the point of view of BG statistical mechanics have been presented. These methods, HCMY and damped potential, are equivalent when used to predict the thermodynamic properties of a selected long-range potential; however, simulation data is required to check their accuracy. A similar treatment can be applied to other long-range potentials.

\section*{Acknowledgements}

We want to acknoweledge CONACyT (proyect-152684) and the University of Guanajuato (DAIP-006/10) for the support in the realization of this work. We thank M. Zachs for improving the presentation of this work.








\newpage

\section*{Figure captions}

\textbf{Figure 1.} Examples of different (repulsive and attractive) long-range potentials, that at long distances decay as $1/r$.\\

\textbf{Figure 2.} The ETS potential with $\alpha^* =1$ (solid line) compared with $-1/x$ potential (dashed line). Long-range behavior is the same for both potentials.\\

\textbf{Figure 3.} Comparison between the ETS potential for $\alpha^*=10$ (solid line) and HCMY potential (dashed line) obtained with the non-linear least squares fitting, with $\kappa_i^*$ and $\epsilon_i^*$ given in Table \ref{table1}.\\

\textbf{Figure 4.} Vapor-liquid phase diagram for a HCAY for $\kappa^*=0.1$. At this value of $\kappa^*$, FMSA and DPT predictions overlap. Diamonds are MC simulation data from Caillol et al.~\cite{Caillol}.\\

\textbf{Figure 5.} Vapor-liquid phase diagram for HCMY approach to the ETS potential. Solid line represents DPT prediction, and squares represent FMSA prediction for HCMY approach with $\epsilon^*_i$ and $\kappa^*_i$, given in Table \ref{table1}.\\

\textbf{Figure 6.} The ETS potential pressure-density plot  within HCMY approach for $T^*=1.2\times10^{8}$ (bottom), $T^*=1.7\times10^{8}$ (middle), both subcritical temperatures,  and a supercritical, $T^*=3.0\times10^8$ (top) for DPT (solid line) and FMSA (dashed line).\\

\textbf{Figure 7.} The ETS potential reduced excess internal energy within HCMY approach for DPT (solid line) and FMSA (dashed line) at two supercritical temperatures $T^*=2\times10^{8}$ (bottom) and $T^*=3\times 10^{8}$ (top).\\

\textbf{Figure 8.} Comparison between the ETS potential for $\alpha^*=10$ (solid line) and the damped potential with damping parameter $\gamma=1.9603\times10^{-4}$ (dashed line).\\

\textbf{Figure 9.} Vapor-liquid phase diagram prediction within DPT of damped approach considering $\gamma=1.9603\times10^{-4}$ and HCMY with $\epsilon_i^*$ and $\kappa_i^*$ given in Table \ref{table1}, the prediction under DPT for both approaches overlap.\\ 

\textbf{Figure 10.} Reduced pressure as a function of reduced density using DPT for damped potential (dashed line) with damping parameter $\gamma=1.9603\times10^{-4}$, and HCMY potential (solid line). Two temperatures are depicted, supercritical  $T^*=3.0\times10^8$ (top) and subcritical $T^*=1.5\times10^8$ (bottom).\\

\textbf{Figure 11.} Reduced excess internal energy  as a function of reduced density using DPT for damped potential approach with a damping parameter $\gamma=1.9603\times10^{-4}$ (dashed line) and HCMY (solid line). Two supercritical isotherms are presented, $T^*=3.0\times10^8$ (bottom) and $T^*=5.0\times10^8$ (top).

\newpage

\section*{Figures}

\begin{figure}[H]
\centering
\includegraphics[width= 9.0cm]{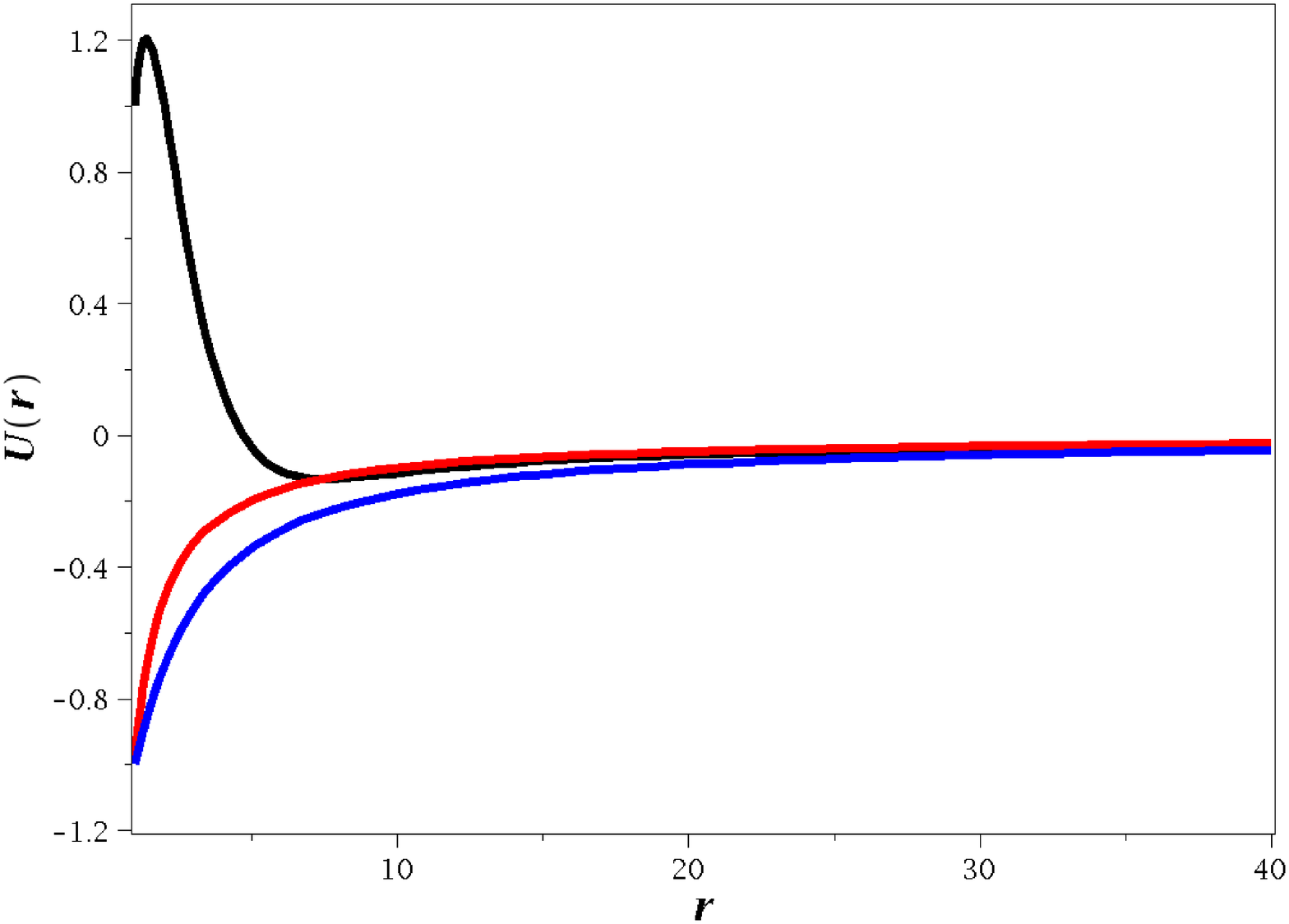} \\
\label{fig1}
\end{figure}
Figure 1.\\

\begin{figure}[H]
\centering
\includegraphics[width= 9.0cm]{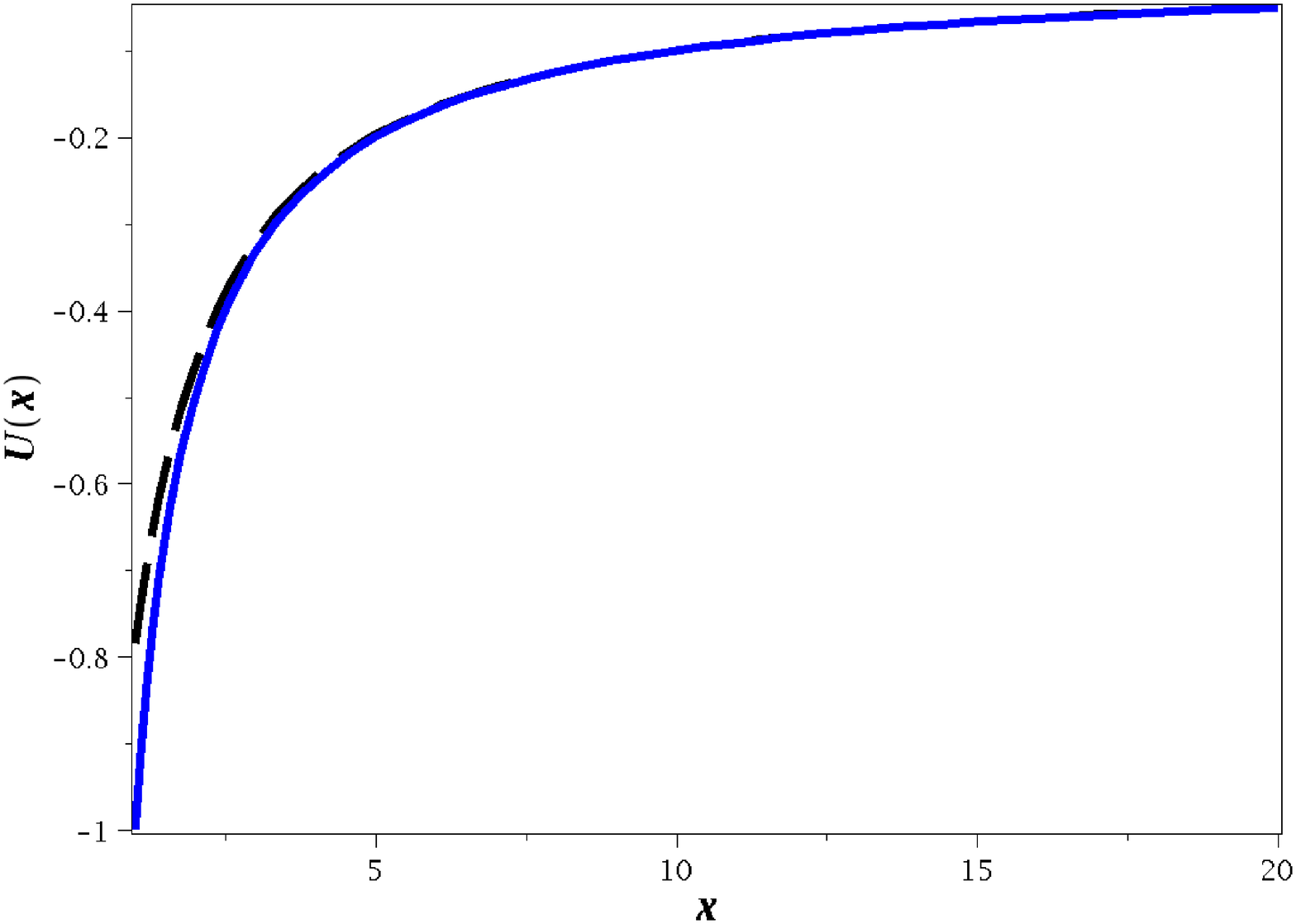} \\ 
\label{fig2}
\end{figure}
Figure 2.\\

\begin{figure}[H]
\centering
\includegraphics[width= 9.0cm]{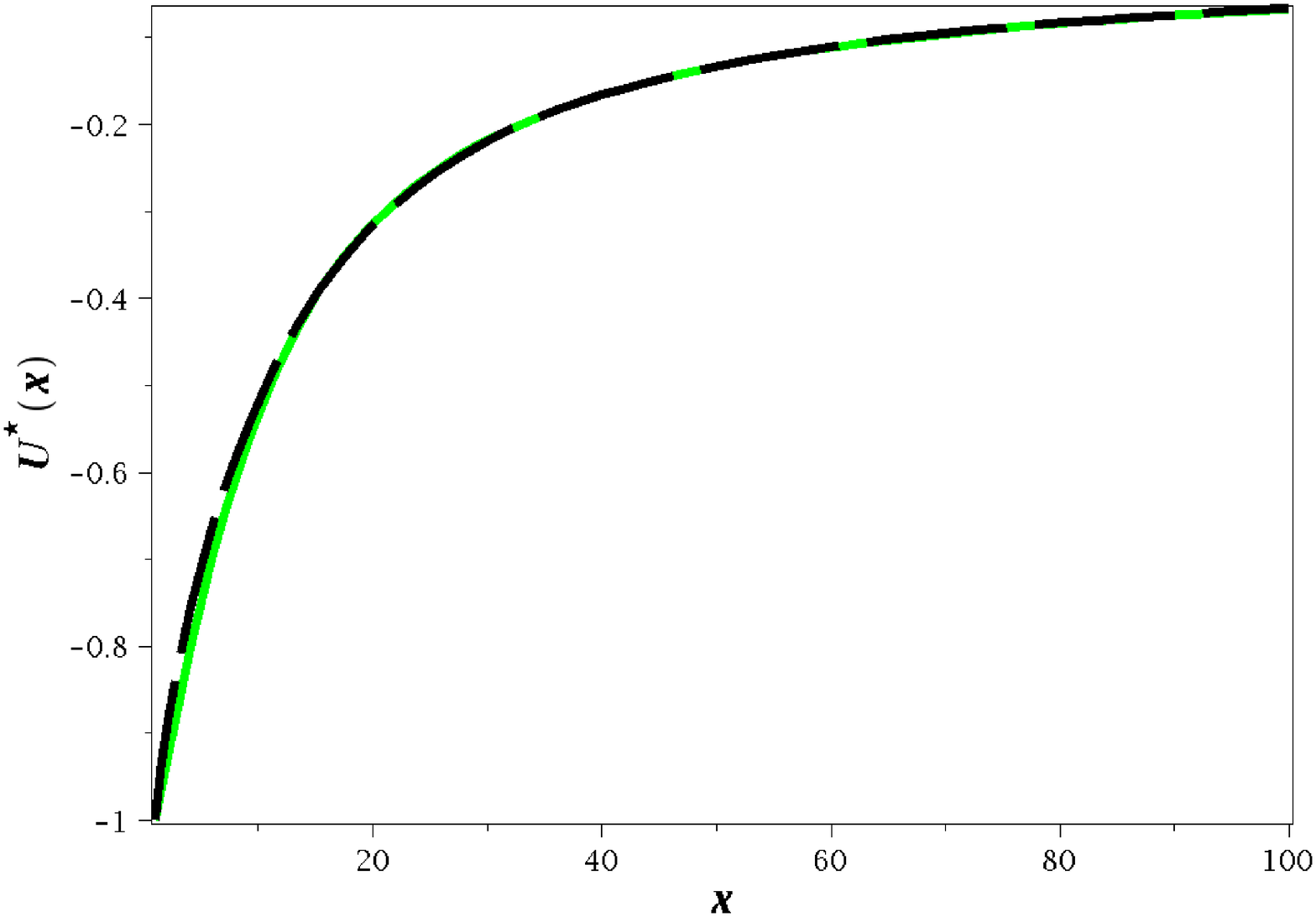} \\ 
\label{fig3}
\end{figure}
Figure 3. \\

\begin{figure}[H]
\centering
\includegraphics[width= 9.0cm]{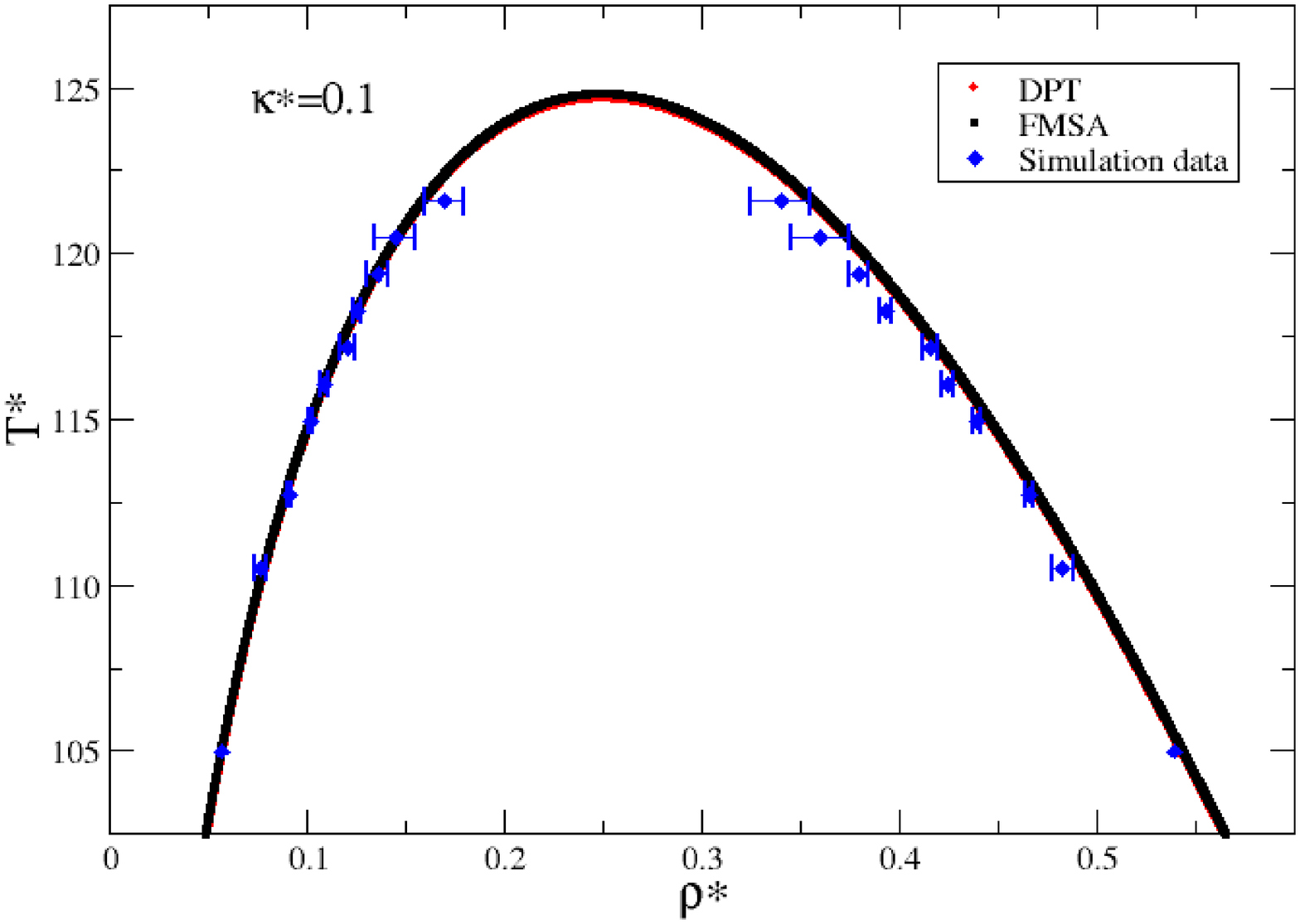} \\ 
\label{fig4}
\end{figure}
Figure 4.\\

\begin{figure}[H]
\centering
\includegraphics[width= 9.0cm]{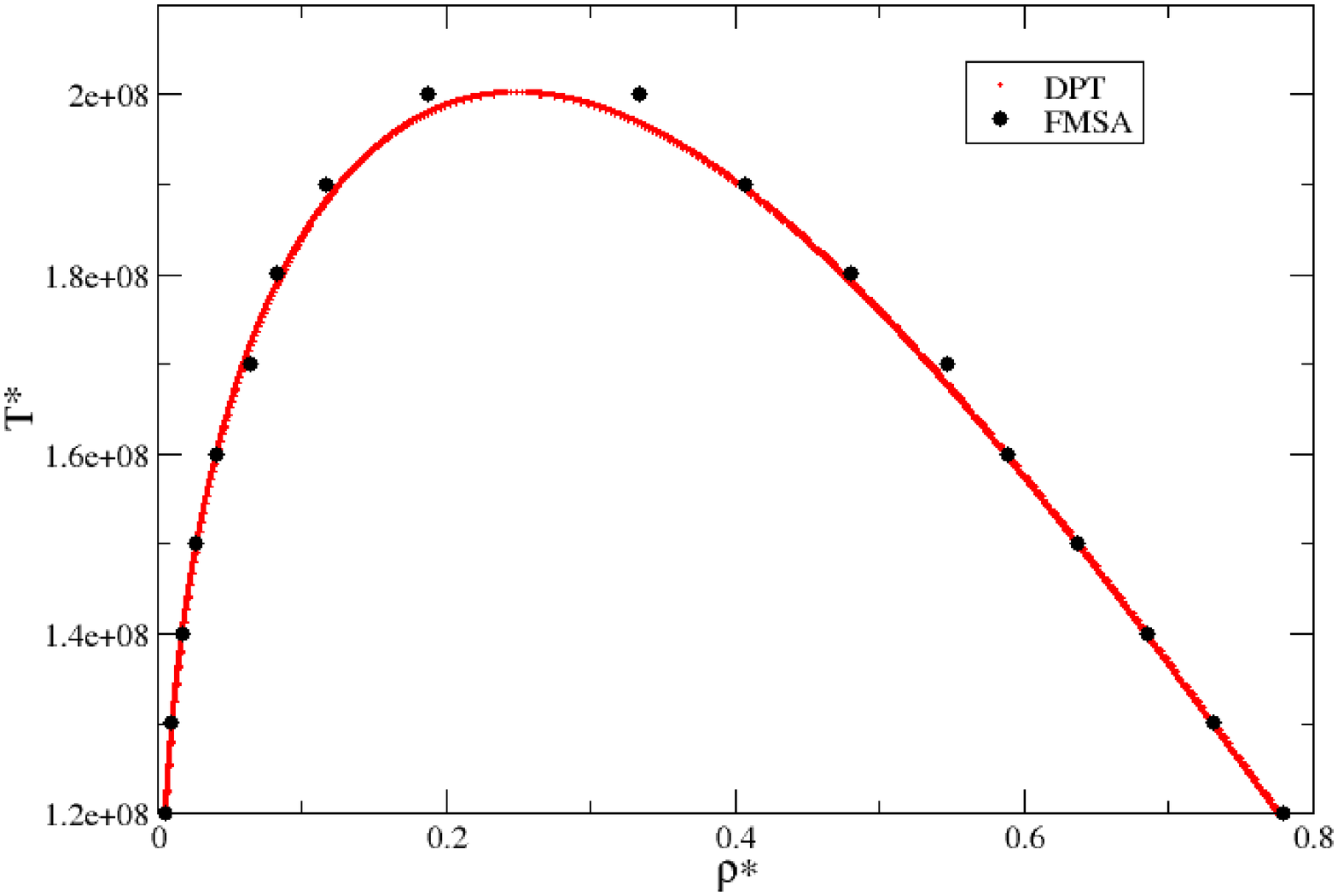} \\ 
\label{fig5}
\end{figure}
Figure 5.\\

\begin{figure}[H]
\centering
\includegraphics[width= 9.0cm]{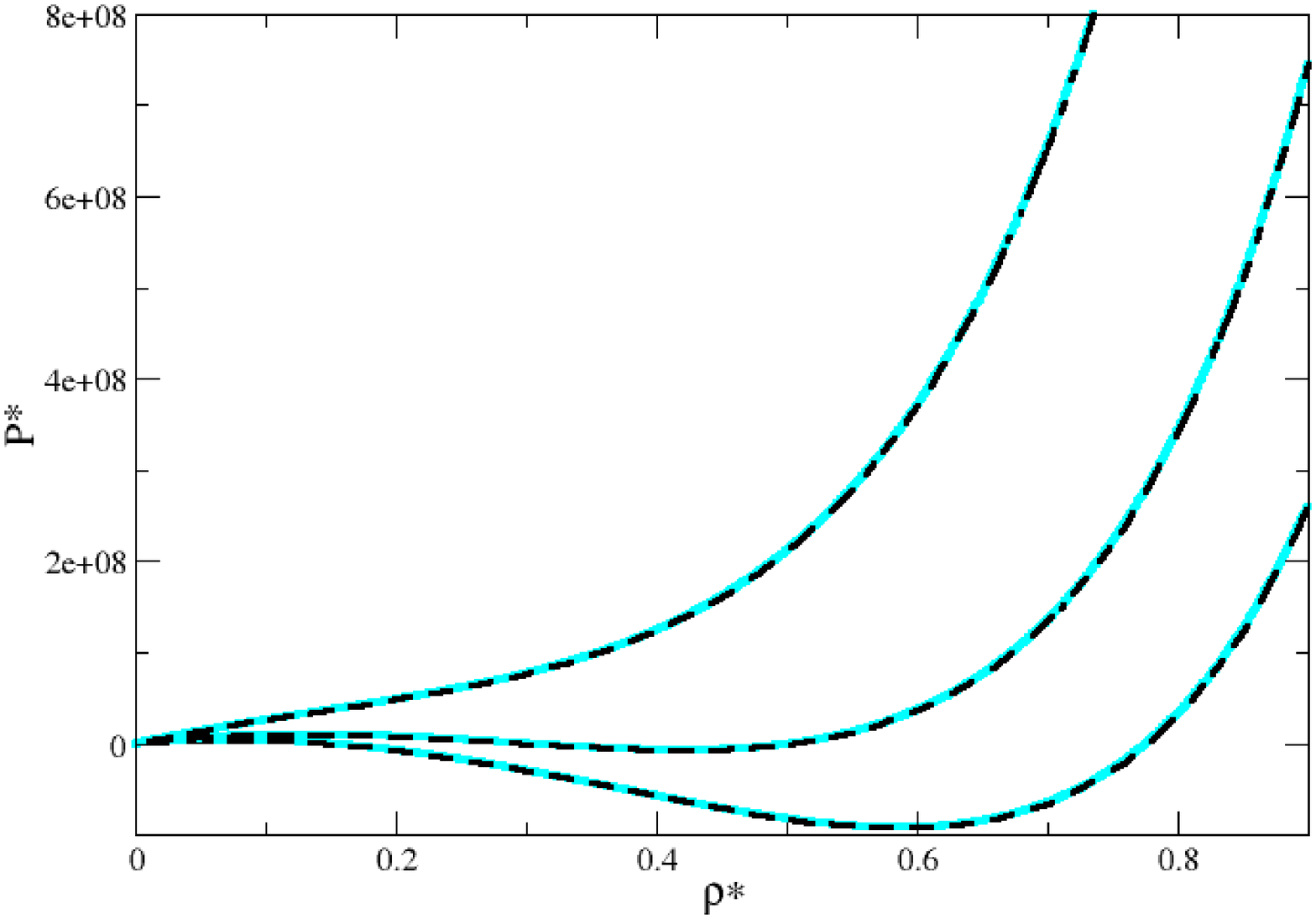} \\  
\label{fig6}
\end{figure}
Figure 6.\\

\begin{figure}[H]
\centering
\includegraphics[width= 9.0cm]{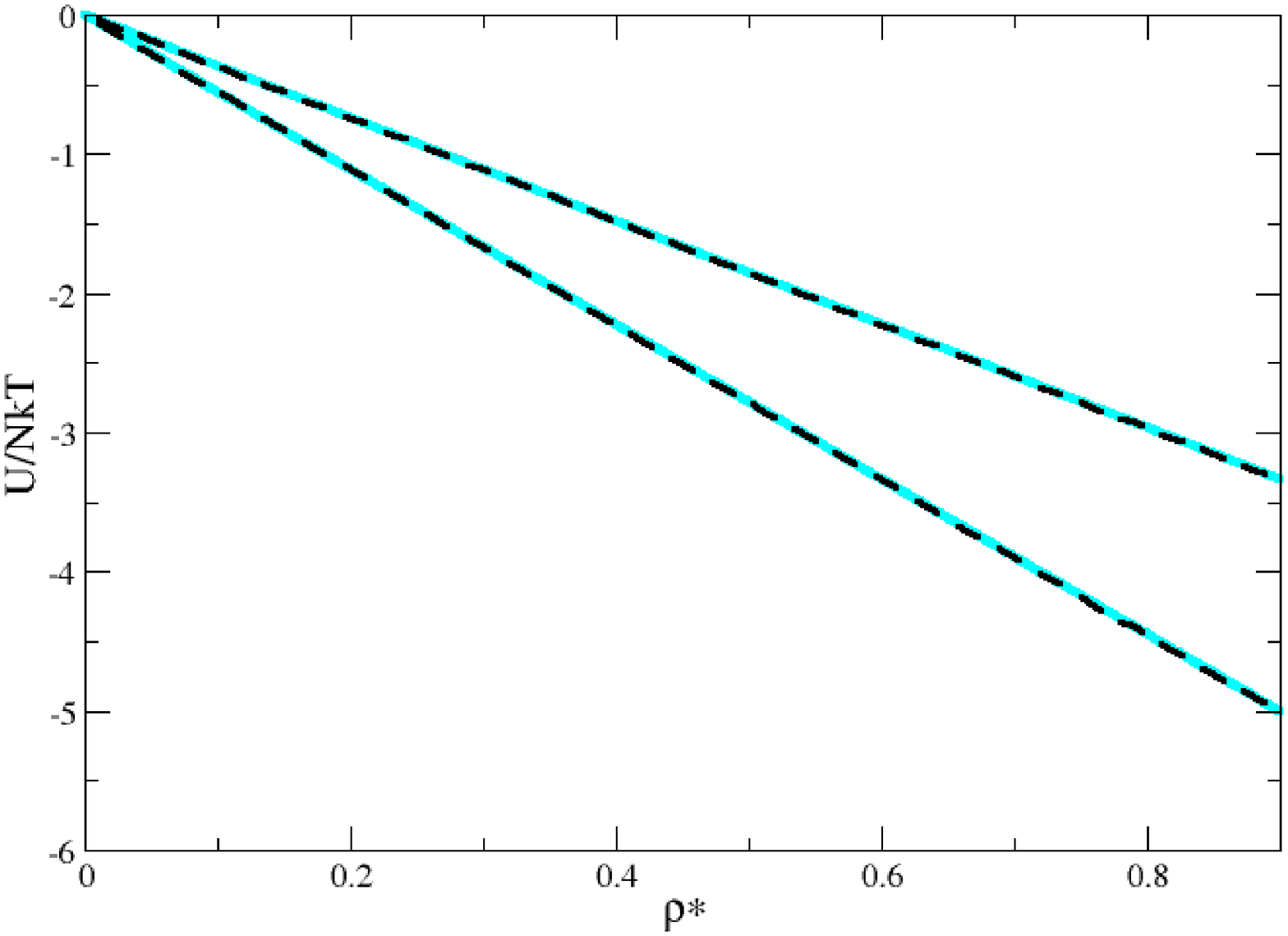} \\  
\label{fig7}
\end{figure}
Figure 7.\\

\begin{figure}[H]
\centering
\includegraphics[width= 9.0cm]{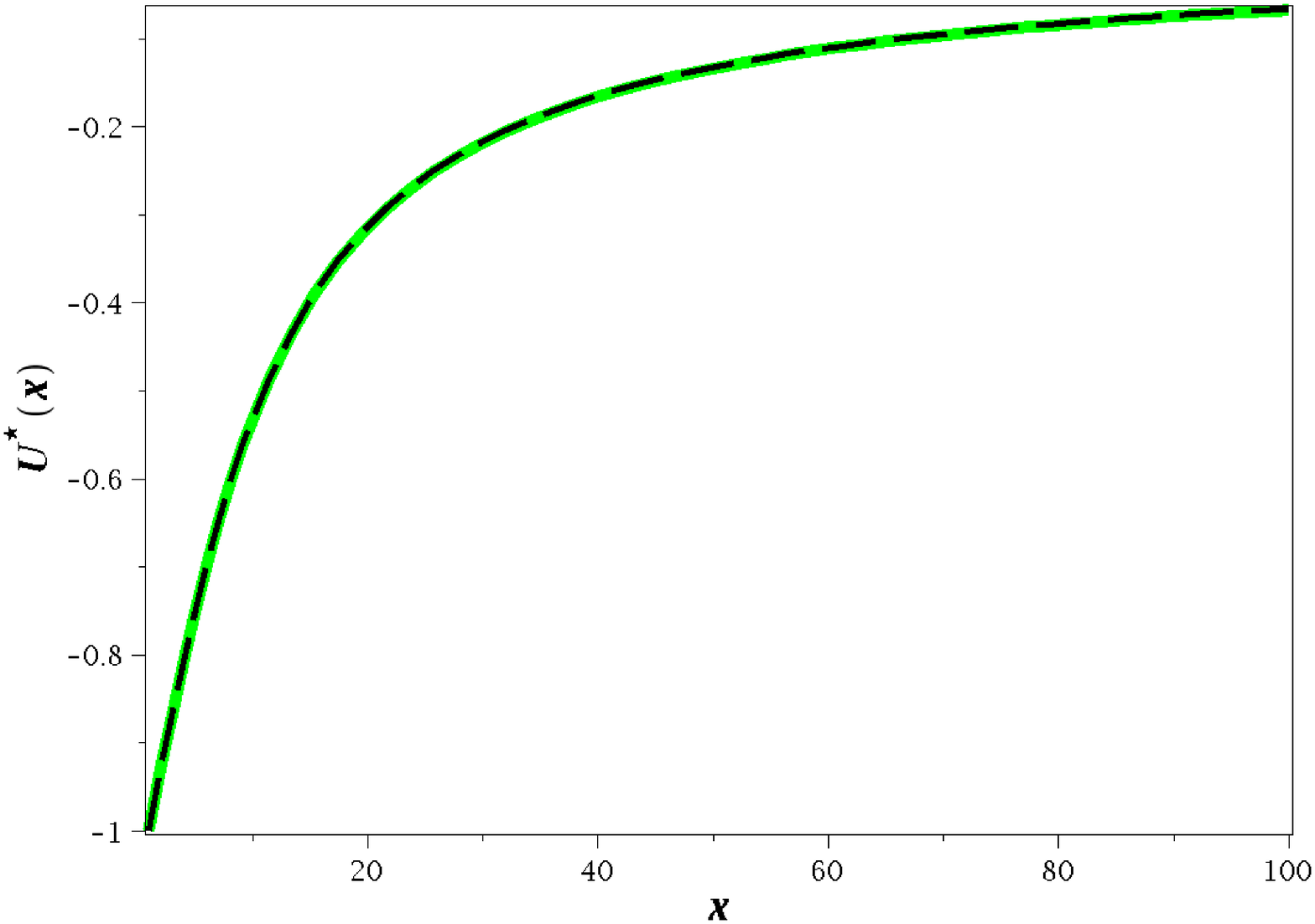} \\ 
\label{fig8}
\end{figure} 
Figure 8.\\

\begin{figure}[H]
\centering
\includegraphics[width= 9.0cm]{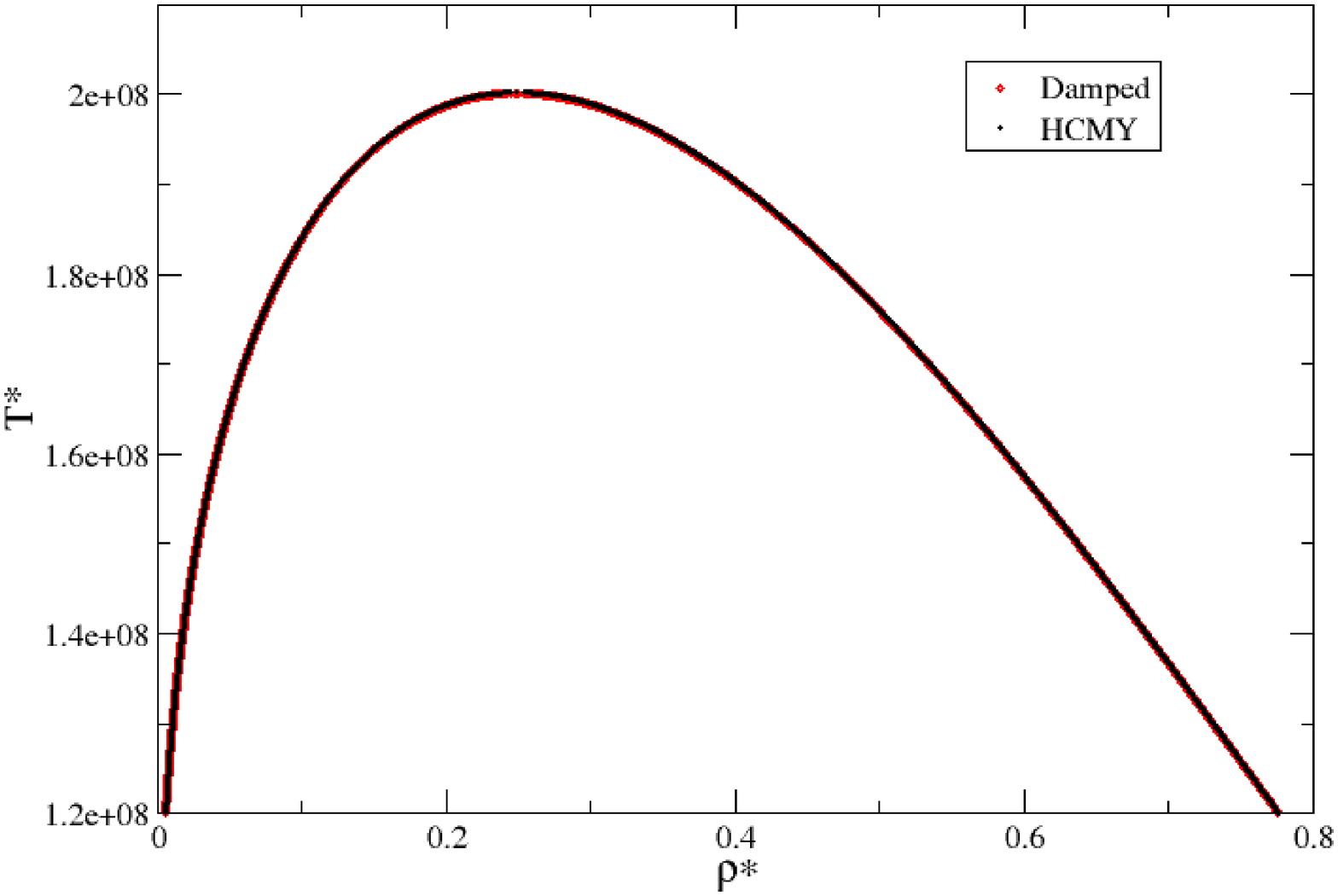} \\ 
\label{fig9}
\end{figure}
Figure 9.\\

\begin{figure}[H]
\centering
\includegraphics[width= 9.0cm]{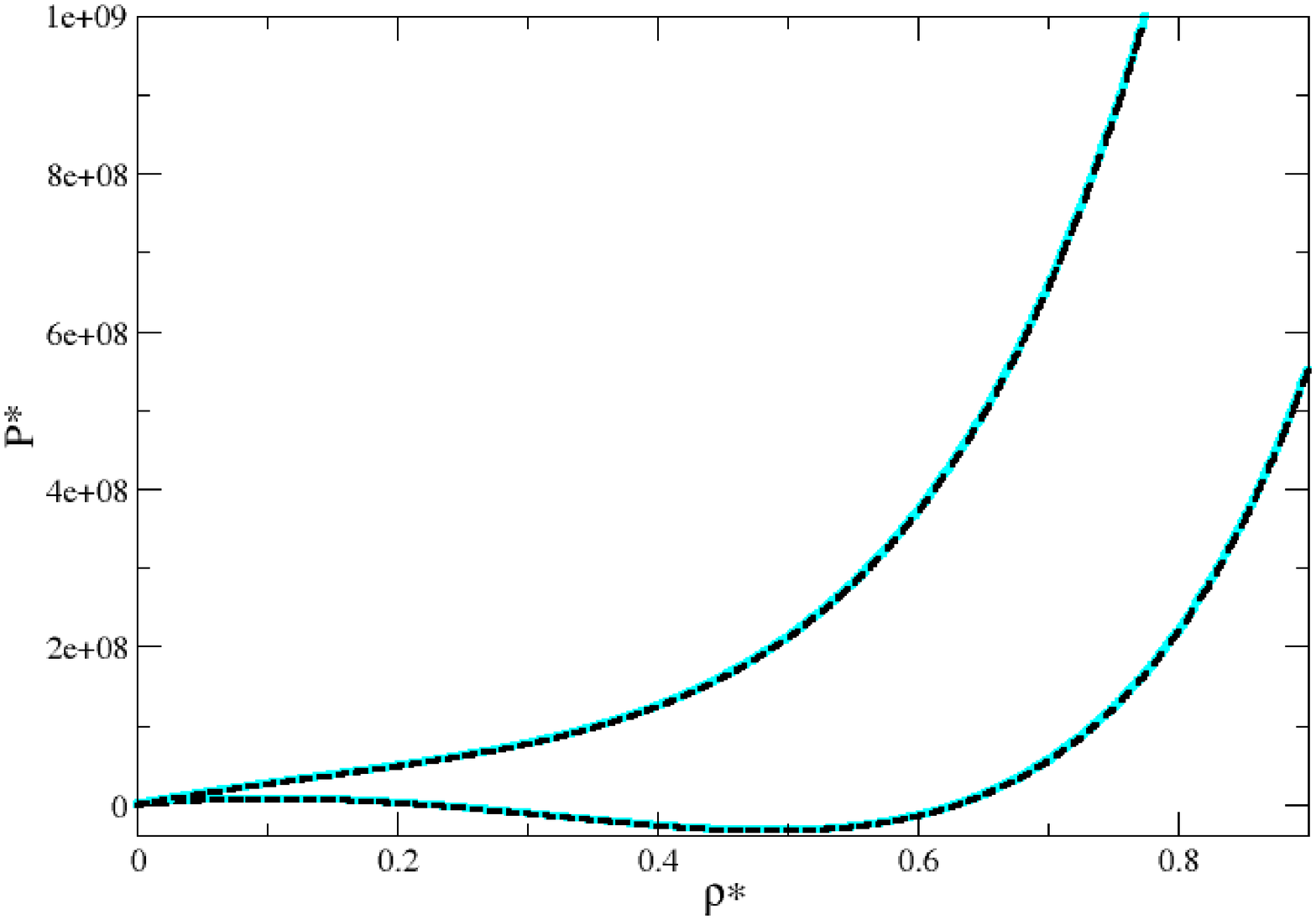} \\ 
\label{fig10}
\end{figure}
Figure 10.\\

\begin{figure}[H]
\centering
\includegraphics[width= 9.0cm]{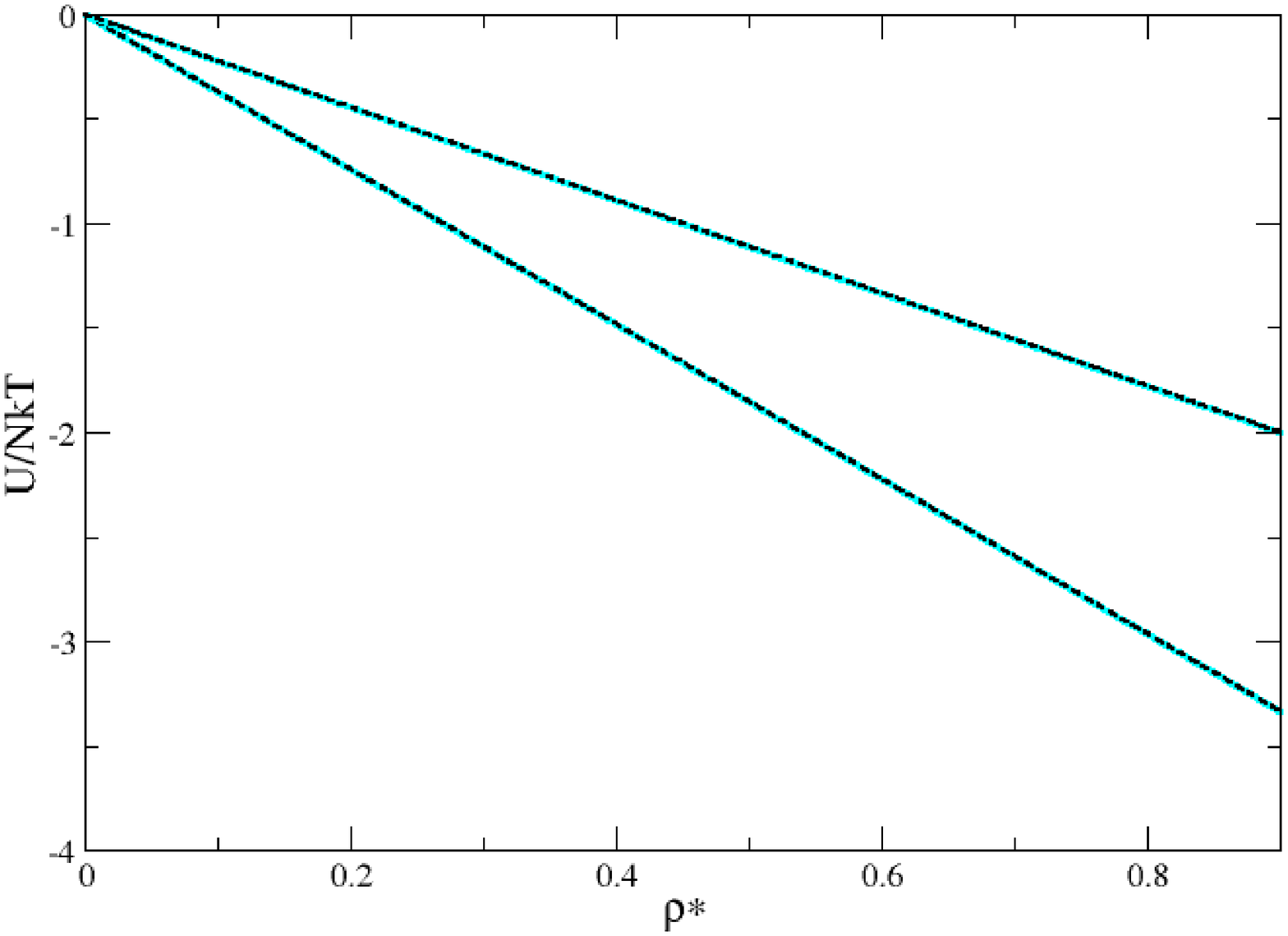} \\ 
\label{fig11}
\end{figure}
Figure 11.\\

\end{document}